%% file: hamann.tex
\documentclass[preprint2]{aastex}
\newcommand \as   {$^{\prime\prime}$}  
\newcommand \cmsq           {\hbox{cm$^{-2}$}}
\newcommand \etal        {{et~al. }}

\newcommand \kms          {\rm{\hbox{km s$^{-1}$}}}
\newcommand \lam          {$\lambda$}
\newcommand \Lya          {\hbox{Ly$\alpha$}}

\newcommand \Msun          {\hbox{M$_{\odot}$}}
\newcommand \pcc           {\hbox{cm$^{-3}$}}

\newcommand \zaz          {{$z_a\kern -1.5pt \approx\kern -1.5pt z_e$}}
\newcommand \zllz         {{$z_a\kern -3pt \ll\kern -3pt z_e$}}

\begin{document}
\title
{\Large\bf High-Resolution Keck Spectra of the Associated\\ 
Absorption Lines in 3C 191}
\bigskip


\author{Frederick W. Hamann\altaffilmark{1}, T.A.
Barlow\altaffilmark{2}, F.C. Chaffee\altaffilmark{3}, C.B.
Foltz\altaffilmark{4}, R.J. Weymann\altaffilmark{5}}

\altaffiltext{1}{Department of Astronomy, University of Florida, 
211 Bryant Space Science Center, Gainesville, FL 32611-2055, 
Internet: hamann@astro.ufl.edu}

\altaffiltext{2}{Infrared Processing and Analysis Center,
California Institute of Technology, 
MS 100-22, 770 South Wilson Ave., Pasadena, CA 91125}

\altaffiltext{3}{California Association for Research in Astronomy, W.H. Keck
Observatory, 65-1120 Mamalahoa Highway, Kamuela, HI 96734}

\altaffiltext{4}{MMT Observatory, University of Arizona, 933 North Cherry
Ave., 
Tucson, AZ 85721-0065}

\altaffiltext{5}{Observatories of the Carnegie Institution of Washington, 
813 Santa Barbara Street, Pasadena, CA 91101-1292}
\begin{abstract} 
Associated absorption lines (AALs) are valuable probes of 
the gaseous environments near quasars. Here we discuss 
high-resolution (6.7 \kms ) spectra of the AALs in the 
radio-loud quasar 3C 191 (redshift $z=1.956$). 
The measured AALs have ionizations ranging 
from \ion{Mg}{1} to \ion{N}{5}, and  
multi-component profiles that are blueshifted by $\sim$400 
to $\sim$1400~\kms\ relative to the quasar's broad emission lines. 
These data yield the following new results.

1) The strengths of excited-state \ion{Si}{2}$^*$ AALs 
indicate a density of $\sim$300 \pcc\ in the Si$^+$ gas. 

2) If the gas is photoionized, this density implies a distance 
of $\sim$28 kpc from the quasar. Several 
arguments suggest that all of the lines form at approximately 
this distance. 

3) The characteristic flow time from the quasar is thus 
$\sim$$3\times 10^7$ yr.

4) Strong \ion{Mg}{1} AALs identify neutral gas 
with very low ionization parameter and high density. 
We estimate $n_H\ga 5\times 10^4$ \pcc\ 
in this region, compared to $\sim$15 \pcc\ where the \ion{N}{5} lines form. 

5) The total column density is $N_{\rm H}\la 4\times 10^{18}$ \cmsq\ in the 
neutral gas and $N_{\rm H}\sim 2\times 10^{20}$ \cmsq\ in the moderately 
ionized regions. These column densities are consistent with 3C 191's  
strong soft X-ray flux and the implied absence of 
soft X-ray absorption.

6) The total mass in the AAL outflow is $M\sim 2\times 10^9$ \Msun , 
assuming a global covering factor (as viewed from the quasar) 
of $\sim$10\% .

7) The absorbing gas only partially covers the background light 
source(s) along our line(s) of sight, requiring absorption in small 
clouds or filaments $<$0.01 pc across. The ratio $N_H/n_H$ 
implies that the clouds have radial (line-of-sight) thicknesses $\la$0.2 pc. 

These properties might characterize a sub-class of AALs that are 
physically related to quasars but form at large distances. 
We propose a model for the absorber in which pockets of dense neutral 
gas are surrounded by larger clouds of generally lower density and 
higher ionization. This outflowing material might be leftover from a 
blowout associated with a nuclear starburst, the onset of quasar 
activity or a past broad absorption line (BAL) wind phase. 

\end{abstract}

\bigskip
\section{Introduction}

Associated absorption lines (AALs) are important diagnostics of the 
gaseous environments of quasars and active galactic nuclei (AGNs). 
The lines are defined empirically as having velocity widths less 
than a few hundred \kms\ and absorption 
redshifts, $z_a$, within a few thousand \kms\ of the quasar's 
emission-line redshift, $z_e$ (Weymann \etal 1979, 
Foltz \etal 1986). The requirement for \zaz\ makes AALs more 
likely to be physically related to the quasars 
than are the other narrow absorption lines (at \zllz ) 
in quasar spectra  (see Rauch 1998 for a review of the unrelated 
\zllz\ systems). The narrow velocity widths further
distinguish AALs from the class of broad absorption 
lines (BALs), whose widths and maximum blueshifted velocities 
both typically exceed 10,000 \kms\ (Weymann \etal 1991). 
BALs clearly 
form in high-velocity winds from quasar engines 
(e.g. Turnshek 1988), 
but AALs can form potentially in a variety of environments 
--- ranging from energetic outflows like the BALs to relatively 
quiescent gas at large galactic or inter-galactic distances 
(see Hamann \& Brandt 2000 for a general review, also 
Tripp \etal 1996, Hamann \etal 1997a, Barlow \& Sargent 1997, 
Barlow, Hamann \& Sargent 1997). More work is needed to locate 
individual AAL absorbers, quantify their kinematic and physical 
properties, and understand the role of the AGNs and/or host 
galaxies in providing the source of their material and kinetic 
energy. 

One interesting property is that, among radio-loud quasars, 
AALs appear more frequently and with greater strength 
in sources with ``steep'' radio spectra and/or 
lobe-dominated radio morphologies (Wills \etal 1995, Barthel \etal 
1997, Richards \etal 2000, also Brotherton \etal 1998 and references 
therein). This weak correlation is usually attributed to an 
orientation effect, whereby AAL regions reside preferentially near 
a disk or torus that is aligned perpendicular to the radio jet 
axis (but see Richards \etal 2000 for an alternative interpretation). 
However, it is not clear if this (possible) AAL geometry 
has its origins on small scales related to the inner black 
hole/accretion disk, or on much larger scales related to the 
host galaxy. It is also not clear if a disk-like geometry 
applies as well to the AALs in radio-quiet sources. 

3C 191 (Q0802+103, $z_e = 1.956$) is a radio-loud 
quasar having both strong AALs (Burbidge, Lynds \& Burbidge 
1966, Stockton \& Lynds 1966, Williams \etal 1975, Anderson 
\etal 1987) and a bipolar, lobe-dominated radio 
structure (Akujor et al. 1994). It therefore follows the 
AAL--radio morphology correlation noted above.
3C 191 also provides a rare opportunity to define the 
distance between the quasar and 
the absorbing gas because several of its AALs arise from 
excited energy states, e.g. \ion{C}{2}$^*$ \lam 1336 and 
\ion{Si}{2}$^*$ \lam 1265,1533 (Bahcall, Sargent \& Schmidt 1967, 
Williams \etal 1975). The strengths of the excited-state lines, 
compared to their resonant counterparts, \ion{C}{2} \lam 1335, 
\ion{Si}{2} \lam 1260,1527, provide measures of the 
gas density needed to populate the upper levels. The density in 
turn constrains the absorber's distance from the quasar, 
with the reasonable assumption that the gas is in 
photoioization equilibrium with the quasar radiation field. 

Williams et al. (1975) already estimated a density of 
$n_e\sim 1000$ \pcc\ and a radial distance of $R\sim 10$ kpc for 
the AAL region in 3C 191. We reobserved 3C 191 with 
higher spectral resolution and wider wavelength coverage 
to 1) obtain more reliable densities as a function 
of velocity, 2) search for \ion{Fe}{2}$^*$ AALs, which might be 
revealing of a much higher density environment (Wampler \etal 1995, 
Halpern \etal 1996), 3) revisit the question of this 
absorber's origin and location, and 4) obtain better constraints 
on the AAL region dynamics, abundances and overall physical 
structure. 

Sections 2 and 3 below describe the observations and 
results. Section 4 provides measurements and analyses of 
the AALs. Section 5 draws further inferences and  
discusses physical models. 
Throughout this paper we define solar abundances by the 
meteoritic results in Grevesse \& Anders (1989), and we 
use atomic transition data from the compilation 
by Verner, Verner \& Ferland (1996).

\section{Observations Data Reductions}

We observed 3C~191 on 3 occasions using the High Resolution Echelle 
Spectrometer (HIRES, Vogt et al. 1994) on the Keck I telescope on 
Mauna 
Kea, Hawaii. On 13 December 1996 we obtained useful  
spectra from $\sim$3850 \AA\ to $\sim$5975 \AA\ in a total of 
12,000 s 
exposure time. On 21 December 1997 we covered the wavelengths 
$\sim$6474 to $\sim$8927 \AA\ in a total of 18,000 s, and on  
1 January 1998 we measured $\sim$3530 \AA\ to $\sim$4935 \AA\ 
in a total of 12,000 s. On each occasion, a 0.86\as\ slit provided 
spectral resolution $\lambda/\Delta\lambda \approx 45,000$ 
or 6.7~\kms , corresponding 
to 3 pixels on the 2048$\times$2048 Tektronix CCD. Also on each 
occasion, we used two settings of the echelle grating and 
cross disperser to achieve continuous wavelength coverage 
across the quoted intervals. There is considerable 
overlap between instrument settings in the short wavelength spectra 
(December 1996 and January 1998), but almost none at longer 
wavelengths (December 1997). Wavelength regions not 
covered by both setups received half the total exposure times listed 
above. 

We use standard techniques and a software package 
called MAKEE\footnote{MAKEE was developed by T.A. Barlow 
specifically for reduction of Keck HIRES data. It is freely 
available on the world wide web at the Keck Observatory home page, 
http://www2.keck.hawaii.edu:3636/} for the initial data reductions 
and spectral extractions. Flat-fielding and wavelength calibrations 
were achieved by measuring internal quartz and Th-Ar arc lamps, 
respectively, during each observing run. The wavelength solutions 
typically have 
root-mean-square uncertainties of $<$0.22 \kms\ 
($<$0.1 CCD pixels). 

We use the IRAF\footnote{IRAF is maintained 
and distributed by the National Optical Astronomy Observatories, 
in cooperation with the National Science Foundation.} software package 
for additional data processing. In particular, we produce normalized 
spectra in each echelle order by fitting the continuum plus broad 
emission lines (BELs) with low-order polynomials. We then average together 
the overlapping echelle orders from all exposures on each night, 
weighting each pixel by the calculated noise (variance). 
There is no evidence for significant variability in the 
AALs measured in common on December 1996 and January 1998.   
We therefore combine these short-wavelength data 
into a single variance-weighted mean spectrum. 

Sharp absorption lines corresponding to the telluric A and B 
bands of O$_2$ overlap with the \ion{Fe}{2} \lam 2587 
and \lam 2600 AALs. 
We attempt to remove these telluric lines by dividing by the 
spectrum of a hot standard star measured the same night at the 
same 
airmass as 3C 191. Those efforts were only moderately 
successful (see Figure 2 below).

Finally, we calibrate the flux using low-resolution spectra 
of 3C 191 and a stellar standard obtained 
at the Lick Observatory on 2 December 1997. The Lick spectra 
were measured through an 8\as\ slit under photometric conditions. 
The final fluxes should have $3\sigma$ uncertainties of $\la$7\%. 

\section{Results and Measurements}

Figure 1 shows part of the combined HIRES spectrum, 
with pixels binned to match (roughly) the plot resolution. 
Figure 2 shows velocity profiles for the strongest AALs. 
The quasar rest frame is defined here and throughout this paper
by the emission-line redshift, $z_e = 1.956$, as measured by 
Tytler \& Fan 
(1992) from \ion{C}{4} \lam 1549, 
\ion{He}{2} \lam 1640, \ion{O}{3}] \lam 1664, and 
\ion{C}{3}] \lam 1909. To improve the signal-to-noise ratio 
across the \ion{N}{5} AALs, Figure 2 plots the linear 
average of the \ion{N}{5} \lam\lam 1239,1243 doublet pair 
(labeled ``\ion{N}{5} average'' in the figure). 

Table 1 lists all of the detected AALs with 
their rest-frame equivalent widths ($W_{\lambda}$) and 
corresponding minimum column densities 
($\log N$). To accommodate some of the line profile 
complexity, we measure these quantities in three 
velocity intervals: $-1400 \leq \Delta v_1 < -1160$ \kms , 
$-1160 \leq \Delta v_2 < -810$ \kms , $-810 \leq \Delta v_3 \leq -400$ 
\kms , plus the total profiles from $-1400$ to $-400$ \kms . 
The equivalent widths listed for ``\ion{N}{5} average'' are 
based on the average profile in this doublet. 
The total equivalent width given for \ion{C}{2}+\ion{C}{2}$^*$ 
applies to the entire blend of \lam 1335+\lam 1336 (Fig. 2). 
For \ion{C}{4} \lam 1548 and \lam 1551, some values of 
$W_{\lambda}$ are omitted because those velocity intervals are 
dominated by the other component of this blended doublet. 
The middle interval ($v_2$) has equal contributions from 
both \ion{C}{4} components. The total \ion{C}{4} equivalent width 
(listed separately as ``\ion{C}{4} total'' in the table) applies 
to the entire blend. The equivalent widths for 
\ion{Fe}{2} \lam 2600 and \ion{Fe}{2} \lam 2587 are based 
on crude linear interpolations across the residual narrow O$_2$ 
absorption lines in these profiles (see Fig. 2 and \S2). 

The uncertainties in the $W_{\lambda}$ results are dominated 
by the (largely) subjective continuum placement. Repeated measurements 
with different continuum levels suggest that the 1$\sigma$ uncertainties 
are $\sim$5\% for lines with $W_{\lambda}\ga 3$ \AA , 
$\sim$10\% for $0.3\la W_{\lambda}\la 3$ \AA , 
and $\sim$20\% for weaker lines. A few weak or blended lines are marked 
with `:' in Table 1, indicating uncertainties $>$20\%.

The column densities listed in Table 1 are discussed in \S4.3 below. 

\section{Analysis of the AAL System}

\subsection{Line Profiles and Kinematics}

Most of the AALs have complex, multi-component velocity 
profiles. These components 
tend to be broader in the stronger (deeper) lines. For example, 
the full width at half minimum (FWHM) of the feature at 
$-$542 \kms\ increases 
from $\sim$13--15 \kms\ in the \ion{Si}{2}, \ion{Al}{2} and 
\ion{Al}{3} lines, to $\sim$30 \kms\ in \ion{Mg}{2}, to $>$100 \kms\ 
in \ion{C}{4}. Other components, e.g. near $-$1000 \kms , blend 
together 
and lose their distinct identities altogether 
in the stronger lines. 
One of the strongest lines, \Lya , has no discrete components 
at all --- only a single broad trough with FWHM $\approx 920$ \kms . 

Overall, the AALs span blueshifted velocities from 
$\sim$400 to $\sim$1400 \kms\ (relative to $z_e=1.956$).
The absolute velocities in the quasar rest frame are uncertain 
by up to a few hundred \kms\ because the BELs, 
used to define $z_e$ (\S3), can themselves be 
either blue or redshifted relative to the quasar (see 
Tytler \& Fan 1992, Marziani \etal 1996). Nonetheless, the measured 
velocity shifts give strong evidence for an outflow from the quasar.

\subsection{Partial Line-of-Sight Coverage}

Some of the strongest AALs, such as \Lya\ and the \ion{C}{4} doublet, 
have flat-bottomed profiles that do not reach 
zero intensity. This profile shape suggests that the transitions are 
very optically thick. 
In principle, the non-zero line intensities could also result from 
many narrow, unresolved line components that only partially overlap 
in velocity. However, such line components cannot be narrower than 
a thermal line width. In an ionized plasma with a nominal temperature 
of $\sim$10$^4$ K, the thermal speeds of 
H and C correspond to FWHMs of $\sim$21 and $\sim$6.5 \kms , respectively. 
These minimum line widths are greater than or equal to the resolution 
of our spectra. Therefore, the lines are fully resolved and the 
flat-bottomed profiles must be caused by  
optically thick absorption {\it plus} unabsorbed flux that fills in the 
bottoms of the troughs. This filling-in can occur if the absorber 
does not fully cover the background light source(s) along our line(s) of 
sight (see Hamann \& Brandt 2000). In general, fully resolved line 
intensities depend on the line-of-sight coverage fraction, $C_f$, as,
\begin{equation}
I_v\ =\ (1-C_f)\,I_o\ +\ C_f\,I_o\,e^{-\tau_v}
\end{equation}
where $0\le C_f\le 1$, $I_o$ is the unabsorbed intensity at a given 
profile velocity $v$, and $I_v$ and $\tau_v$ are the measured intensity 
and true optical depth at that velocity. For $\tau_v\gg 1$, the coverage 
fraction simply equals the depth of the line below the continuum, 
such that,
\begin{equation}
C_f\ \ge\ 1 - {{I_v}\over{I_o}}
\end{equation}
If the flat-bottom
lines in 3C 191 are, in fact, optically thick, the 
residual intensities in their cores imply $C_f\approx 96$\% for the 
\ion{C}{4} doublet, $\sim$93\% for \ion{Si}{4}, and $\sim$88\% 
for \Lya . 

Further evidence for $C_f < 1$ appears in Figure 3, which 
compares the measured profiles in 
\ion{Mg}{2} \lam 2804, \ion{Al}{3} \lam 1863 and \ion{Fe}{2} 
\lam 2344 to predictions based on their stronger multiplet 
counterparts \ion{Mg}{2} \lam 2796, \ion{Al}{3} \lam 1855 and 
\ion{Fe}{2} \lam 2383. The predictions follow by 1) 
deriving the {\it apparent} optical depths, 
$\tau_v^a \equiv -\ln (I_v/I_o)$, in the 
stronger lines 
(i.e. assuming $C_f=1$ in Eqn. 1), 2) calculating $\tau_v^a$ in the weaker 
transitions by scaling by the relative $gf\lambda$ values, 
and 3) using those calculated optical depths to predict $I_v$ 
in the weaker lines (see also Barlow \& Sargent 1997). 

The weaker predicted 
compared to observed lines in Figure 3 require $C_f<1$. 
The average intensities from $-$1025 to $-$860 \kms\ (near 
the line centers) differ between the observed and predicted 
profiles at significance levels of 
$\sim$16$\sigma$ in the \ion{Mg}{2} doublet, 
$\sim$14$\sigma$ in \ion{Al}{3}, and $\sim$2.7$\sigma$ in the 
\ion{Fe}{2} pair (where $\sigma$ represents 1 standard 
deviation 
in the photon counting statistics). The 
average coverage fractions in this velocity interval 
are $\langle C_f\rangle = 0.84\pm 0.01$ in \ion{Mg}{2}, 
$0.60\pm 0.02$ in \ion{Al}{3}, and 0.57 +0.13/$-$0.09 in 
\ion{Fe}{2} (see Hamann \etal 1997a, Barlow \& Sargent 1997, 
Petitjean \& Srianand 1999 for explicit equations). 
In the deepest part of the lines, the coverage fractions 
are larger. For example, notice that the core of the 
\ion{Al}{3} \lam 1855 line dips well below $1- \langle C_f\rangle
\approx 0.4$ in Figure 2. The line intensities in the narrower 
interval $-$950 to $-$915 \kms\ imply
$\langle C_f\rangle = 0.89\pm 0.02$ in \ion{Mg}{2}, 
$0.79\pm 0.02$ in \ion{Al}{3}, and 0.68 +0.22/$-$0.9 in 
\ion{Fe}{2}. Similar variations in $C_f$ across AAL 
profiles have been noted previously in other quasars (Barlow \etal 
1997).

\subsection{Column Densities}

The column densities in Table 1 are lower limits derived 
from the apparent line optical depths as follows,
\begin{equation}
N\ \geq\ {{m_ec}\over{\pi e^2 f\lambda_o}}\,\int\tau_v^a\ dv
\end{equation}
where $\lambda_o$ is the rest wavelength of the line 
(see Savage \& Sembach 1991). The equality in Equation 3 holds 
only if $C_f=1$ and the line absorption is fully resolved. 
Otherwise the result is a lower limit. 
To avoid spurious spikes in the $\tau_v^a$ distributions
(e.g. where $I_v\la 0$ in ``noisy'' lines), we first 
smooth the spectra several times with a binomial algorithm and 
then impose an arbitrary upper limit of $\tau_v^a \leq 6$ at each 
data point. Tests show that the integrated results are not sensitive 
to the value of this upper limit, as long as it is not much above 
$\sim$10. Table 1 lists the resulting minimum column densities in each 
of the velocity intervals $\Delta v_1$, $\Delta v_2$ and $\Delta v_3$, 
and in the total profiles (see \S3). 

The total column density listed for \ion{C}{2}+\ion{C}{2}$^*$ 
comes from treating this blended pair as a single line with 1/2 the 
average 
\ion{C}{2}~\lam 1335 +\ion{C}{2}$^*$~\lam 1336 $gf$ value. 
The column density given as ``\ion{C}{4} total'' is the sum of 
contributions from $\Delta v_1$ in \lam 1548 plus $\Delta v_2$ and 
$\Delta v_3$ in \lam 1551. This estimate should approximate the 
total \ion{C}{4} column density. Finally, the results 
for ``\ion{N}{5} average'' are derived from the average profile 
of this doublet (Fig. 2, \S3) assuming an average $f$ value. 

The evidence for partial coverage in \S 4.2 implies that the 
column densities listed in Table 1 are, in fact, just lower limits. 
These limits could be orders 
of magnitude too low for strong, flat-bottomed lines like \Lya , 
\ion{Si}{3} \lam 1206, \ion{Si}{4} \lam\lam 1394,1403  
and \ion{C}{4} \lam\lam 1548,1551 (see Figure 2). For weak and 
intermediate lines, we can estimate the true column densities  
by adopting nominal covering fractions in our derivations of 
the $\tau_v^a$ (using Equation 1). We adopt $C_f = 0.85$ for 
the \ion{Mg}{2} doublet and $C_f = 0.7$ for the other lines 
(\S4.2). The resulting corrected column densities are listed 
for the total profiles under the heading ``Tot($C_f<1$)'' 
in Table 1. 

\subsection{Ionization}

The metallic AALs range in ionization from \ion{Mg}{1} \lam 2853 
to \ion{N}{5} \lam 1239,1243. 
This range of ions requires a range of distinct ionization 
zones. In particular, \ion{Mg}{1} indicates a  
neutral gas component that cannot survive in the highly-ionized 
\ion{N}{5} region. The upper limit to the ionization 
is unknown because important lines of higher ionization species, 
such as \ion{O}{6}~\lam\lam 1032,1038, are outside our wavelength 
coverage. The degree of neutrality in the \ion{Mg}{1} zone 
can be inferred from theoretical models. 

We examine  
theoretical clouds in photoionization equilibrium with  
the quasar radiation field using the computer 
code CLOUDY (version 90.05, Ferland \etal 1998). The 
ionizing spectrum has the form $L_{\nu}\propto \nu^{\alpha}$, 
where $\alpha = -1.6$ for $\nu \geq \nu_{LL}$ and $\alpha = -0.7$ 
for $\nu < \nu_{LL}$ (where $\nu_{LL}$ is the frequency at 
the \ion{H}{1} Lyman limit). The spectral slope at $\nu < \nu_{LL}$ 
approximates the measured rest-frame UV spectrum from 
$\sim$1200 \AA\ to $\sim$2300 \AA\ (Fig. 1). 
The slope in the Lyman continuum is not known, 
but a spectral steepening to $\alpha\sim -1.6$ at $\nu\ga \nu_{LL}$ 
is consistent with the best observations of other luminous 
quasars (Hamann, Netzer \& Shields 2000, Zheng \etal 1997, 
Laor \etal 1997).  The ionization parameter, $U$, is defined as the 
dimensionless ratio of hydrogen particle to hydrogen-ionizing 
photon densities at the illuminated face of the clouds,
\begin{equation}
U\ \equiv\ {{1}\over{4\pi\, c\, R^2\, n_H}}\, \int^{\infty}_{\nu_{LL}} 
{{L_{\nu}}\over{h\nu}}\, d\nu
\end{equation}
where $L_{\nu}$ is the luminosity density of the quasar spectrum, 
$R$ is the distance between the absorber and the quasar, and 
$n_H$ is the total hydrogen density (in \ion{H}{1} + \ion{H}{2}). 

A critical question is whether or not the absorber contains 
an \ion{H}{2}--\ion{H}{1} recombination front. If there is 
such a front, the absorption of Lyman continuum 
photons might allow low-ionization species like \ion{Si}{2}, 
\ion{Al}{2} and \ion{Fe}{2} to exist physically adjacent to 
the high-ionization \ion{C}{4} and \ion{N}{5} region, 
i.e. at essentially the 
same $R$. Individual clouds (at a given $R$ and $n_H$) 
could have high $U$ at their illuminated surface and still 
host low-ionization species behind the front. If, on the other hand, 
there is no ionization front, the absorber would be optically thin 
in the Lyman continuum ($\tau_{LC}\la 1$) 
and there could be no (significant) gradient in the ionization 
level due to the absorber's own opacity. The observed 
range of ionizations would then require a variety of clouds 
having different $U$ values and 
therefore very different $n_H$ and/or $R$ (Eqn. 4). 

A key constraint comes from \ion{Mg}{1} (Arav \etal 2000). Its 
ionization potential, 7.65 eV (corresponding to $\lambda \sim 1620$ 
\AA ), is well below the \ion{H}{1} Lyman limit of 13.6 eV (912 \AA ). 
Therefore \ion{Mg}{1} cannot be shielded 
by the presence of an \ion{H}{2}--\ion{H}{1} recombination front. 
Explicit CLOUDY calculations show that clouds with $U$ high 
enough to amply support \ion{N}{5} near their illuminated surface 
cannot also support significant amounts of \ion{Mg}{1} behind 
an \ion{H}{2}--\ion{H}{1} front. Essentially all of the magnesium 
behind the front is in the form of \ion{Mg}{2} because the 
strong photon flux at $912\la\lambda\la 1620$ \AA\ photoionizes 
\ion{Mg}{1}, while the absence of photons at $\lambda\la 912$ \AA\ 
prevents the formation of \ion{Mg}{3}. (Al, Si 
and Fe behave similarly.) The observed ratio of 
\ion{Mg}{1}/\ion{Mg}{2} column densities, $\sim$0.05 (Table 1), 
is several orders of magnitude larger than the predicted value 
everywhere in these model clouds. Further calculations 
with $\log U = -2$ to $-$3 similarly underpredict $N$(MgI)/$N$(MgII) 
by factors of $\sim$10--20 compared to Table 1. 

The only way to match the observed $N$(MgI)/$N$(MgII) ratio in 
high-$U$ clouds is to invoke an additional opacity source to 
suppress the flux between 912 and 1620 \AA\ by a factor 
of at least ten. Bound-free absorption by neutral 
metals cannot provide this opacity because the column densities 
implied by the neutral lines are too low 
(e.g. $\log N($\ion{Mg}{1}$)({\rm cm}^{-2}) \approx 13.2$); 
moreover, we do not observe absorption edges 
at the appropriate wavelengths (e.g. at 1620 \AA\ for \ion{Mg}{1}). 
Molecules, such as H$_2$, also 
cannot be important opacity sources because there are no edges 
at their dissociation energies (e.g. at 4.48 eV for H$_2$). 
The only viable possibility is dust inside the absorbing region. 
However, the required large reduction in UV flux should produce 
significant reddening or perhaps a broad $\sim$2200 \AA\ absorption 
feature (e.g. Cardelli, Clayton \& Mathis 1989, Fall \& Pei 1989, 
Brotherton \etal 1998, Meurer, Heckman \& Calzetti 1999 and 
references therein). There is no evidence for 
these dust signatures in 3C 191 (compare Fig. 1 and the UV 
slope estimated above with O'Brien, 
Gondhalekar \& Wilson 1988 and Zheng \etal 1997). 

The large observed $N$(MgI)/$N$(MgII) ratio therefore cannot be 
attributed to radiative shielding; rather, it requires neutral 
gas with very low ionization parameter. \ion{Mg}{1} and \ion{N}{5} 
together imply that the overall absorber contains a wide range of 
$U$ values and therefore a wide range in $n_H$ and/or $R$. 

Figure 4 shows theoretical ionization fractions for clouds with 
$\tau_{LC}\la 1$ and different $U$ (see Hamann 1997 
for more details). Results 
for $\log U\la -5$ are not shown because they are increasingly 
sensitive to uncertain factors, such as the input volume density, 
the derived temperature, and the details of molecule formation. 
Nonetheless, it seems clear that the  
measured ratio of \ion{Mg}{1}/\ion{Mg}{2} column densities 
requires predominantly neutral gas with $\log U \la -5$. 
In contrast, the strong \ion{N}{5} AALs 
identify a highly-ionized region where $\log U \sim -1.5$ is 
more representative. Intermediate ionizations are also present; 
for example, the ratio of \ion{Al}{2}/\ion{Al}{3} column 
densities indicates $\log U \approx -2.8$ in Figure 4. 

Overall there 
is a trend for increasing column densities in higher ionization 
species. In particular, the high-ionization lines are all saturated 
(Figure 2, \S4.3), and the inferred ratios of \ion{Si}{2}/\ion{Si}{3}, 
\ion{Al}{2}/\ion{Al}{3} and \ion{C}{2}/\ion{C}{4} column densities 
are all above unity (Table 1). Clearly, there is much more 
highly-ionized compared to neutral gas in the overall AAL region. 

\subsection{Total Column Density and X-ray Absorption}

We can quantify these statements by estimating the total column 
density, 
$N_H$ (in \ion{H}{1}+\ion{H}{2}), in different regions. 
For example, an upper limit on $N_H\approx N$(\ion{H}{1}) in the 
neutral gas follows from the measured column density in 
\ion{Mg}{1}. If the Mg/H abundance ratio 
is roughly solar, i.e. $\log ({\rm Mg/H})\approx -4.4$ 
(which would be consistent with AALs in other 
quasars --- Petitjean, Rauch \& Carswell 1994, Hamann 1997, 
Hamann \& Ferland 1999), and the \ion{Mg}{1} ionization fraction 
is conservatively $\log f({\rm MgI}) \ga -1$ in the neutral gas, 
then $N_{\rm H}\approx N$(\ion{H}{1}) should be $\la$5.4 dex larger 
than $N$(\ion{Mg}{1}) in Table 1. Therefore, we expect 
$\log N_{\rm H}({\rm cm}^{-2}) \la 18.6$ in the neutral absorber. 

For the higher ions, we assume the ionization fractions are near 
their peaks in Figure 4 to derive conservatively low estimates 
of $N_{\rm H}$ in their line-forming regions. Again assuming solar 
metal-to-hydrogen abundance ratios, we find 
$\log N_{\rm H} ({\rm cm}^{-2}) \approx 19.5$ and 19.4 based on the 
measured \ion{Si}{2} and \ion{Al}{2} column densities, respectively, 
$\log N_{\rm H} ({\rm cm}^{-2}) \approx 20.3$ based on \ion{Al}{3}, 
and $>$19.6 based on a minimum \ion{Si}{3}+\ion{Si}{4} column density. 
The total column density in more highly ionized regions could be 
even larger, given the saturated absorption in 
\ion{Si}{4}, \ion{C}{4} and \ion{N}{5} (\S3). 
In any case, $N_{\rm H}$ is at least 1.7 dex larger  
at moderate to high ionizations compared to the 
neutral gas. 

The total column density of $\log N_{\rm H} ({\rm cm}^{-2}) 
\approx 20.3$ derived above corresponds to a continuum optical depth 
of $\la$5\% at 1 keV (with the maximum value obtaining for a neutral 
absorber, Morrison \& McCammon 1983). This prediction based on the 
AALs is consistent with the observed normal ratio of soft X-ray to 
UV continuum fluxes in 3C 191 (Wilkes \etal 1994). 
There is evidently no significant soft X-ray absorption  
in this source. This situation contrasts with the extreme  
X-ray absorption in BAL quasars, and with the strong to moderate 
X-ray absorption found in other AGNs with significant AALs 
(Green \& Mathur 1996, Gallagher \etal 1999, 
Crenshaw \etal 1999, Brandt, Laor \& Wills 2000).

\subsection{Excited-State AALs: Space Density and Radial Distance}

Several excited-state AALs of \ion{C}{2}$^*$ and \ion{Si}{2}$^*$ 
are present (Table 1 and Figure 2). 
These features arise from ground $^2P^o$ 
multiplets that should behave approximately as 
2-level atoms --- with level populations controlled by collisional 
processes and forbidden 
radiative decays (see Morris \etal 1986). Bahcall \& Wolf 
(1968) showed that electron impacts are more important than proton 
collisions for temperatures below $\sim$20,000 K, which should be 
appropriate for 
the C$^{+}$ and Si$^+$ regions in photoionized plasmas.  
Absorption lines from the excited states, e.g. 
\ion{Si}{2}$^*$ \lam 1533 and \ion{C}{2}$^*$ \lam 1336, compared 
to the resonance transitions, \ion{Si}{2} \lam 1527 and \ion{C}{2} 
\lam 1335, provide direct measures of the level populations and 
thus the electron density. The density is given by,
\begin{equation}
n_e\ \approx\ n_{cr}\,\left({{2\,N_{lo}}\over{N_{up}}}\,-\,1\right)^{-1} 
\end{equation}
where $N_{lo}$ and $N_{up}$ are the column densities 
measured from the ground and excited-state lines, respectively, and 
$n_{cr}$ is the critical electron density of the upper state (at which 
collisional deexcitation equals radiative decays). 
Note that $N_{lo}/N_{up}$ must be $\ge$0.5. The critical 
density scales with the electron temperature roughly as $T^{1/2}$. 
For $T=8000$K, we derive $n_{cr}\approx 34$ \pcc\ for the 
C$^{+}$ upper state and $n_{cr}\approx 1580$ \pcc\ for Si$^+$ 
(using collision strengths from Osterbrock 1989 and radiative decay 
rates from Mendoza 1983 and Galavis, Mendoza \& Zeippen 1998). 

Inspection of the similar \ion{Si}{2} \lam 1527 and 
\ion{Si}{2}$^*$ \lam 1533 profiles indicates a similar 
density at all velocities. In the main AAL components, between 
$-1160$ and $-810$ \kms\ ($\Delta v_2$ in Table 1), the column 
densities inferred from these lines imply $n_e\approx 510$ \pcc\ 
if $C_f = 1$ or $n_e\approx 300$ \pcc\ if $C_f\approx 0.7$. The narrow 
absorption component at $-$542 \kms\ in these transitions similarly 
indicates $n_e\approx 395$ \pcc\ if $C_f \approx 1$ or $n_e\approx 300$ 
\pcc\ if $C_f = 0.7$. 

The measured \ion{C}{2} \lam 1335 and 
\ion{C}{2}$^*$ 
\lam 1336 lines are noisy and may be saturated. 
Nonetheless, their relative strengths suggest 
densities above the critical value of $\sim$34 \pcc\ ---
consistent with the \ion{Si}{2}/\ion{Si}{2}$^*$ results. 
We do not detect any \ion{Fe}{2}$^*$ lines, even 
though several strong transitions lie within our spectral coverage. 
An explicit calculation (kindly performed by G. Ferland 
using CLOUDY) shows that for $n_e\approx 300$ \pcc\ and 
$T\approx 8000$ K the strongest measurable 
\ion{Fe}{2}$^*$ line, \lam 2349, should be $\ga$40 times 
weaker than \ion{Fe}{2} \lam 2383 and therefore below our detection 
threshold. We therefore
adopt $n_e\approx 300$ \pcc\ as being 
representative of the low-ionization gas. 

Assuming the gas is photoionized, we can combine this density with 
$U$ from \S4.4 to estimate the distance, $R$, between the 
absorber and the quasar (Eqn. 4). We approximate $\L_{\nu}$ 
in the Lyman continuum by a power law of the form, 
\begin{equation}
L_{\nu}\ =\ L_{LL}\, \left({{\nu}\over{\nu_{LL}}}\right)^{\alpha}
\end{equation}
where $L_{LL}$ is the luminosity density at the Lyman limit. 
Equation 4 then yields,
\begin{equation}
R\ =\ \left({{-L_{LL}}\over{4\pi\, c\, h\, n_H\, U\, \alpha}}\right)^{1/2}
\end{equation}
for $\alpha < 0$. 
We estimate $L_{LL}\approx 3.6\times 10^{31}$ ergs s$^{-1}$ Hz$^{-1}$ 
in the quasar rest frame by extrapolation from the fluxes in Figure 1 
(assuming a cosmology with $H_o = 65$ \kms\ Mpc$^{-1}$ 
and $q_o = \Omega_M/2 - \Omega_{\Lambda} = 0$; Carroll, 
Press \& Turner 1992). We again adopt $\alpha = -1.6$ in the Lyman 
continuum (\S4.4). 

If we choose $n_{\rm H}\approx n_e\approx 300$ \pcc\ and a 
nominal value of $\log U \approx -2.8$ 
in the \ion{Si}{2} zone (based on the \ion{Al}{2}/\ion{Al}{3} 
column densities, \S4.4), the radial distance of that zone 
should be,
\begin{equation}
R\ \approx\ 28 {\rm ~kpc}
\end{equation} 
This result is not sensitive to the assumed spectral shape 
because the photon energies that control the \ion{Si}{2} ionization 
are also most important in the definition of $U$ (i.e. near the 
\ion{H}{1} Lyman limit, Eqn. 4). The value of $U$ presents a 
much larger uncertainty. Williams \etal (1975) derived a smaller 
distance of $R\approx 10$ kpc partly because they combined 
a density based on \ion{Si}{2}/\ion{Si}{2}$^*$ with a higher $U$ 
value based on \ion{Si}{4}, \ion{C}{4} and \ion{N}{5}. That 
derivation, therefore, implicitly assumed that the \ion{Si}{2} 
gas is radiatively shielded downstream from the \ion{C}{4}, 
etc. region. We have assumed that shielding is {\it not} important 
in the \ion{Si}{2} zone, based on the \ion{Mg}{1} strength (\S4.4 
and \S5.1 item 3).

\subsection{Abundance Ratios}

We cannot derive the metallicity of the AAL gas 
because the only measured H line, \Lya ,  is severely saturated. 
The only possibility is to use some of the weaker lines, e.g. of 
Mg, Al, Si and Fe, to estimate the relative abundance ratios in 
these elements. The logarithmic abundance ratio of any two elements 
$a$ and 
$b$ can be written, 
\begin{equation}
\left[{a\over b}\right]\ = \ 
\log\left({{N(a_i)}\over{N(b_j)}}\right)\ +\
\log\left({{f(b_j)}\over{f(a_i)}}\right)\ +\
\log\left({b\over a}\right)_{\odot} 
\end{equation}
where $(b/a)_{\odot}$ is the solar abundance ratio, 
and $N$ and $f$ are respectively the column densities and ionization 
fractions of element $a$ in ion state $i$, etc. 

Figure 5 shows theoretical values of the normalized 
ionization corrections, defined here as $IC\equiv 
\log(f(b_j)/f(a_i)) + \log (b/a)_{\odot}$, based on the 
calculations in \S4.4 (Figure 4). Figure 5 shows that $IC$ for 
\ion{Mg}{2}/\ion{Fe}{2} and \ion{Si}{2}/\ion{Fe}{2} 
have minimum values near zero (in the log), 
which provide firm lower limits on the Fe/Si and Fe/Mg 
abundances (see Hamann 1997 and Hamann \& Ferland 1999 for more 
discussion). With column densities from Table 1, the corresponding 
abundance limits are [Fe/Mg] $\ga\ -0.3$ and [Fe/Si] $\ga\ -0.9$. 
The lowest ratios would obtain only if $-5.2\la\log U \la -4.5$, 
but that would imply Al/Mg and Al/Si abundances several 
times above solar (Table 1 and Fig. 5). If the actual $U$ values 
are instead in the range $-3.4 \la\ \log U\ \la\ -2.8$ 
(consistent with our estimate of 
$\log U \sim -2.8$ based on \ion{Al}{2}/\ion{Al}{3}, \S4.4), 
then we would derive roughly solar ratios among all of these 
elements. 

Given the uncertainties in $U$, and the fact that different 
line forming regions can have different $U$ values (\S4.4), 
we conclude only that the relative metal abundances are 
broadly consistent with solar, with uncertainties of factors 
of several. 

\section{Discussion}

\subsection{Implications for the Structure and Dynamics}

The results in \S3 and \S4 lead to following important 
conclusions regarding the AAL environment. 

1) We have already noted that the absorber has a complex 
velocity structure 
that appears qualitatively similar in all lines and ions (\S4.1). 
Stronger AALs, e.g. of higher ionization, are more smoothly
distributed in velocity, but it seems clear that all of the lines 
trace the same overall physical structure. We conclude that 
the radial distance $R\approx 28$ kpc derived for 
the Si$^+$ zone (\S4.6) should be roughly characteristic of the 
entire AAL region (see also item 4 below). 

2) The AAL gas appears to be outflowing from the quasar 
at velocities from $\sim$400 to $\sim$1400 \kms\ (\S4.1). If the 
gas is not accelerating (or decelerating), then the time scale 
for this outflow reaching radius $R$ is
\begin{equation}
t\ \approx\ 3\times 10^7\ \left({{R}\over{28\, {\rm kpc}}}\right) 
\left({{1000\, {\rm km\, s}^{-1}}\over{v}}\right)~~{\rm yr}
\end{equation}
where $v$ is a characteristic velocity. The overall AAL region 
will acquire a radial thickness during this expansion because 
different gas components move at different speeds. In 3C 191, 
the radial thickness, $\Delta R$, could be comparable to 
the radius, $R\approx 28$ kpc, because the velocity dispersion 
implied by the line widths, $\Delta v \approx 1000$ \kms , is 
similar to the average flow speed. 

3) The significant presence of \ion{Mg}{1}, compared to \ion{Mg}{2}, 
identifies a neutral gas component whose 
survival cannot be attributed to radiative shielding 
downstream from an \ion{H}{2}--\ion{H}{1} recombination front 
(\S4.4). The \ion{Mg}{1} region must have a low 
ionization parameter and therefore a high density. If we adopt 
$n_H\approx 300$ \pcc\ and $\log U\approx -2.8$ for the 
\ion{Si}{2} region (\S4.6), then simple scaling based on 
$\log U\la -5$ in the neutral 
gas (\S4.4) implies $n_H\ga 5\times 10^4$ \pcc\ in that region. 
In contrast, $\log U\sim -1.5$ in the \ion{N}{5} zone 
indicates $n_H\sim 15$ \pcc\ there. 

4) The \ion{Si}{2} \lam 1527 and \ion{Si}{2}$^*$ \lam 1533 
line profiles are surprisingly similar (Fig. 2) given the 
wide range of densities present. The nearly 
constant line ratio indicates a nearly constant density 
(within a factor of $\sim$2) at 
all velocities. This result could be caused by a selection 
bias: the absorber may have a wide range of densities at 
each velocity, but the specific ionization requirements of 
Si$^+$ might persistently lead to just a narrow range in $U$ 
and therefore $n_{\rm H}$ controlling the \ion{Si}{2} and 
\ion{Si}{2}$^*$ line strengths. In any case, the similar 
$n_{\rm H}$ values inferred for 
different \ion{Si}{2} velocity components support 
the argument (item 1 above) that the AAL region does not span 
a wide range in radial distance from the quasar. 

5) Partial line-of-sight coverage of the quasar emission sources 
(\S4.2) implies that the AAL clouds have characteristic sizes 
similar to or smaller than the projected 
area of the emitters. The \Lya\ and \ion{C}{4} 
absorption lines sit atop strong BELs (Fig. 2); they 
might fully cover the continuum source while partly covering the 
larger BEL region. The characteristic size of these absorbers might 
therefore be as large as 0.1 -- 1 pc (i.e. the size of the BEL 
region, e.g. Peterson 1993, Kaspi \etal 2000). However, the partial 
coverage inferred from the 
\ion{Mg}{2}, \ion{Al}{3} and \ion{Fe}{2} lines must involve the 
continuum source, which requires absorber size scales $<$0.01 pc 
(for standard accretion disk models of the continuum emission, 
Netzer 1992). 

6) Another constraint on the size scales comes 
from the ratio $N_H/n_H$. 
If the absorbing gas completely fills the volume it 
encompasses, then the radial thickness of the entire absorber 
would be of order,
\begin{equation}
\Delta R\ \approx\ 0.2\, \left({{N_H}\over{2\times 10^{20}\, 
{\rm cm}^{-2}}}\right) \left({{300\, {\rm cm}^{-3}}\over{n_H}}\right) 
~~{\rm pc}
\end{equation}
using parameters from \S4.5 and \S4.6. 
However, if the absorber is composed of discrete clouds that fill only 
part of the encompassed volume, then the overall AAL region could 
have a much greater radial thickness (item 2 above) while the 
individuals clouds are small compared to $\Delta R$ in Equation 11.  

7) The mass of the AAL region, $M$, depends on its radial distance, 
total column density and 
{\it global} covering factor, $Q\equiv \Omega /4\pi$ (where 
$\Omega$ is the solid angle subtended by the absorber as ``seen'' 
from the central quasar). The value of $Q$ is not known, but it is 
not likely to exceed the detection frequency of AALs 
among radio-loud quasars\footnote{The detection frequency sets an 
approximate upper limit on $Q$ because some fraction of the 
systems counted as AALs will have a different physical origin 
than the absorber in 3C 191 (\S1).}, e.g. $\sim$30\% (G. Richards, private 
communication). If we let $Q_{0.1}$ represent the covering factor  
relative to $Q=0.1$, the total mass is given by
$$M\ \approx\ 2\times 10^{9} ~ Q_{0.1} \left({{\mu_H}\over{1.4}}\right) 
\left({{N_H}\over{2\times 10^{20}\, {\rm cm}^{-2}}}\right) 
$$
\begin{equation}
\times\ \left({{R}\over{28\, {\rm kpc}}}\right)^2 ~{\rm M}_{\odot}
\end{equation}
where $\mu_H$ is the mean molecular weight per H particle. 
The total kinetic energy in the AAL outflow is therefore,
\begin{equation}
K\ \approx\ 2\times 10^{58}\,\left({{M}\over{3\times 10^9\, \Msun}}\right) 
\left({{v}\over{1000\, {\rm km\, s}^{-1}}}\right)^2 ~ {\rm ergs}
\end{equation}
For comparison, the much faster ($v\sim 10^4$ \kms ) 
but smaller-scale ($R\sim 0.1$ pc) 
BAL outflows observed in other quasars are believed to contain 
total masses (at any instant) of $\sim$1--10 \Msun\ for $Q_{0.1}\sim 1$ 
to 3. Over a quasar's lifetime, say 10$^8$ yr (Haehnelt, 
Natarajan \& Rees 1998), a BAL wind could eject a total of 
$\sim$10$^7$--10$^8$ \Msun\ with $K \approx\ 10^{58}$--10$^{59}$ 
ergs (Hamann \& Brandt 2000). 

8) Emission lines from the AAL gas could be quite strong, 
depending on the actual values of $U$ or $N_{\rm H}$. For example, 
we calculated emission line strengths for a photoionized plasma 
consistent with the \ion{Si}{2} AAL region described above, namely, 
having solar abundances, an incident spectrum as defined in \S4.4, 
and the following physical paramters: 
$R\approx 28$ kpc, $\log U\approx -2.8$, $N_{\rm H} = 2\times 10^{20}$ 
\cmsq , and $n_H\approx 300$ \pcc . 
The strongest predicted emission lines within our 
wavelength coverage for 3C 191 are \Lya\ and \ion{Mg}{2} 
\lam 2799, with  rest-frame equivalent widths of $\sim$44$Q_{0.1}$ \AA\ 
and $\sim$4$Q_{0.1}$ \AA . These results should be compared 
to the rest-frame equivalent widths of the measured BELs, 
e.g. $\sim$120 \AA\ in \Lya\ and $\sim$17 \AA\ in \ion{Mg}{2} 
(as estimated without the superposed AALs). 
The predicted emission lines could therefore 
be present but ``hidden'' in the measured BELs. Clearly, however, 
$Q_{0.1}\sim 1$ is close to an upper limit by this test. 
Another constraint is that $Q_{0.1}>1$ could lead to too 
much line emission 
``filling in'' the bottoms of the AAL troughs (depending on the 
global line-of-sight velocity distribution of the emitting gas; 
see Hamann, Korista \& Morris 1993 and Hamann \& Korista 1996 
for similar arguments related to BALs). 
Future observations at longer wavelengths might 
provide more stringent upper limits on $Q_{0.1}$. In particular, 
our calculations predict that the strongest lines 
in the rest-frame near-UV/visible should be H$\alpha$, H$\beta$, 
the \ion{O}{2} \lam 3727 doublet, and \ion{O}{3} \lam 5007, with 
rest equivalent widths of $\sim$$55Q_{0.1}$ \AA , $\sim$11$Q_{0.1}$ \AA , 
$\sim$22$Q_{0.1}$ \AA , and $\sim$88$Q_{0.1}$ \AA , respectively. 

\subsection{Toward a Physical Model}

Figure 6 shows a highly schematic model of the AAL region, wherein 
pockets of dense neutral gas are surrounded by a diffuse, 
spatially distributed medium of generally higher ionization. 
The diffuse clouds contain most of the total column density 
(\S4.5). Their greater size and/or greater numbers 
lead to more complete coverage in both velocity and 
projected area (for example, in the \ion{C}{4} doublet), 
compared to the lower-ionization gas 
(e.g. \ion{Mg}{1} \lam 2853). 

Note that the extended regions cannot have exclusively 
high ionization levels because the measured line-of-sight 
coverage fractions do not correlate simply with ionization. 
For example, low ionization lines can have either high 
(e.g. \ion{C}{2} \lam 1334, \ion{Si}{2} \lam 1260) or low 
(\ion{Fe}{2}) coverage fractions (\S4.2). 
The amount of coverage in both space and velocity must depend 
at least partly on the line's oscillator strength. In other words, 
the coverage fraction scales with the line optical depth. 
Stronger (more optically thick) lines have greater contributions 
from the diffuse extended gas, resulting in greater coverage, 
whereas weak lines sample mainly the 
higher column density material in more compact regions. 

A major concern with any cloud model is the 
cloud survival. The lifetime of a cloud without pressure 
confinement is of order the sound-crossing time. For a 
nominal temperature of $10^4$ K 
and a maximum cloud size of $\Delta R \la\ 0.2$ pc (Eqn. 11), 
the cloud survival time, $\la$$2\times 10^4$ yr, is much less 
than the characteristic flow time, $t\approx\ 3\times 10^7$ yr 
(Eqn 10). Therefore pressure confinement appears necessary. 
The problem of understanding this confinement has plagued 
cloud models of both the BAL and BEL regions of AGNs. 
Possible solutions include external pressure from 
a magnetic field or a surrounding hot, low-density 
(and transparent) plasma (Weymann, Turnshek \& Christiansen 1985, 
Arav, Li \& Begelman 1994, Emmering, Blandford \& Shlosman 1992, 
DeKool 1997, Feldmeier \etal 1997). 

The key remaining question is, what provides the source of 
material and kinetic energy for the AAL outflow? 
The flow time ($\sim$$3\times 10^7$ yr) is comparable to 
predicted quasar lifetimes (Terlevich \& Boyle 1993, 
Haehnelt et al. 1998). It therefore seems likely that the AAL gas 
originated much nearer the quasar, perhaps coincident with 
the onset of quasar activity.

We have already noted (\S5.1, item 7) that the kinetic energy 
in this AAL region is comparable to the typical energy 
in BAL winds (integrated over a quasar's lifetime). 
Therefore, quasars are capable of driving winds with 
this total energy. In addition, most current 
models of BAL winds have them preferentially located near the 
plane of the accretion disk (Murray \& Chiang 1995, 
Emmering \etal 1992, Wills, Brandt \& Laor 1999 and references 
therein). This geometry is reminiscent of the equatorial structure 
(tentatively) inferred for AAL regions in radio-loud quasars (\S1). 

However, it is unlikely 
that the AAL outflow in 3C 191 is simply an extended remnant 
of a BAL wind because 1) the terminal velocity of a BAL wind should 
be of order 10$^4$ \kms\ instead of 1000 \kms , and 2) the total 
mass in AAL gas is at least an order of magnitude larger than 
expected for BAL winds (item 7 in \S5.1). To produce the 
observed AALs, a high-velocity BAL-like wind would have to 
be decelerated by interaction with ambient galactic material 
and then, probably, entrain some of that material (to add mass) 
along the way. 

An alternative possibility is that the AALs form in 
gas that was expelled by 
stellar processes, e.g. in a galactic ``superwind'' as observed in 
low-redshift starburst galaxies (Heckman, Armus \& Miley 1990, 
Heckman \etal 2000). The sizes, masses, velocities, etc. 
inferred for superwinds in luminous starbursts are consistent 
with our estimates of these quantities in 3C 191. The superwinds 
also contain cool dense clouds (giving rise to \ion{Na}{1} 
absorption lines) embedded in a hot ($\sim$10$^7$ K) X-ray emitting 
plasma (see also Heckman \etal 1996 and references therein). 
If a superwind model does apply to 3C 191, the 
characteristic flow time of the AAL gas ($\sim$$3\times 10^7$ yr) 
might represent the time elapsed since the starburst episode. 

No matter what scenario accounts for the AALs in 3C 191, it 
is important to keep in mind that AALs in different objects 
can probe very different physical phenomena. For example, 
AALs in other quasars often have higher blueshifted velocities 
than those in 3C 191. Some narrow absorption lines have 
blueshifts above the arbitrary 5000 \kms\ AAL threshold, 
even though there is strong evidence for their being  intrinsic 
to the quasar environments (Hamann \etal 1997a and 1997b, 
Barlow \& Sargent 1997, Barlow, Hamann \& Sargent 1997, 
Richards 2000). A galactic superwind certainly cannot explain 
these  high-velocity absorbers. Most AAL systems also do not 
have low-ionization lines like 3C 191 (e.g. Junkkarinen, 
Hewitt \& Burbidge 1991, Hamann 1997). It is possible 
that low ionization AALs, which allow us to locate the absorber 
via \ion{Si}{2}$^*$, select in favor of large absorber--quasar 
distances. In particular, all of the known AAL systems with these 
excited-state lines have distances $\ga$10 kpc 
(e.g. Barlow \etal 1997, Tripp, Lu \& Savage 1996, 
Morris \etal 1986, Sargent, Boksenberg \& Young 1982). 
Other AALs are known to form much closer 
to the quasars, possibly within a few pc in 
outflows similar to the BALs (Hamann \etal 1997a and 1997b, 
Barlow \& Sargent 1997, Barlow \etal 1997). 

Given this diversity, it is interesting to note that 3C~191 
does not follow the trend identified by Brandt \etal (2000) 
for small X-ray to UV continuum flux ratios accompanying 
strong \ion{C}{4} absorption equivalent widths (\S4.5). 
That correlation nominally points to a relationship between 
the strength of the AALs and the strength of continuous (bound-free) 
absorption in X-rays. BAL quasars are at one extreme in this 
relationship --- having both strong UV lines and strong absorption 
in X-rays (see also Green \& Mathur 1996, Gallagher \etal 1999). 
3C 191 might contain a different 
class of absorber (e.g. much farther from the active nucleus) 
than the majority of sources discussed by Brandt \etal (2000). 

\acknowledgments

We are grateful to Tom Bida and Bob Goodrich for their 
help with the Keck observations, and Tony Misch for 
assistance at the Lick Observatory. We thank Gary Ferland 
for his continued support and distribution of the CLOUDY 
software, and, in particular, for providing 
the Fe$^+$ level population results discussed in \S4.6.  
The work of FH was funded in part by a NASA 
grant, NAG 5-3234, and an NSF Career Award, 
AST-9984040. CBF acknowledges the support of NSF grant 
AST 98-03072
\bigskip

\include{table1}

\parskip=0pt
\leftskip=0.2in
\parindent=-0.2in
\section*{\bf References}
Akujor, C.E., L\"udke, E., Browne, 
I.W.A., Leahy, J.P., Garrington, S.T., Jackson, N., \& Thomasson, P. 
1994, \aaps, 105, 247

Anderson, S.F., Weymann, R.J., Foltz, C.B. \& Chaffee, F.H. 1987, 
\aj, 94, 278

Arav, N., Li, Z.-Y., \& Begelman, M.C. 1994, \apj, 432, 62

Arav, N., Brotherton, M.S., Becker, R.H., Gregg, M.D., White, R.L., 
Price, T., \& Hack, W. 2000, preprint (astro-ph/0008259)

Bahcall, J.N., Sargent, W.L.W., \& Schmidt, M. 1967, \apj, 149, L11

Bahcall, J.N., \& Wolf, R.A. 1968, \apj, 152, 701

Barlow, T. A., Hamann, F., \& Sargent, W. L. W. 1997, in 
Mass Ejection From AGN, eds. R. Weymann, I. Shlosman, and N. Arav, 
ASP Conf. Series, 128, 13

Barlow, T. A., \& Sargent, W. L. W. 1997, \aj, 113, 136

Barthel, P., Tytler, D.R., \& Vestergaard, M. 1997,  in 
Mass Ejection From AGN, eds. R. Weymann, I. Shlosman, and N. Arav, 
ASP Conf. Series, 128, 48

Brandt, W.N., Laor, A., \& Wills, B.J. 2000, \apj, 528, 637

Brotherton, M.S., Wills, B.J., Dey, A., van Bruegel, W., \& Antonucci, 
R. 1998, \apj, 501, 110

Burbidge, E.M., Lynds, C.R., \& Burbidge, G.R. 1966, \apj, 144, 447

Cardelli, J.A., Clayton, G.C., \& Mathis, J.S. 1989, \apj, 345, 245

Carroll, S.M., Press, W.H., \& Turner, E.L. 1992, \araa, 30, 499

Crenshaw, D.M., Kraemer, S.B., Boggess, A., Stephen, P., 
Mushotzky, R.F., \& Wu, C.-C. 1999, \apj, 516, 750

de Kool, M. 1997, in 
Mass Ejection From AGN, eds. R. Weymann, I. Shlosman, and N. Arav, 
ASP Conf. Series, 128, 233

Emmering, R.T., Blandford, R.D., \& Shlosman, I. 1992, \apj, 385, 460

Fall, S.M., \& Pei, Y.C. 1989, \apj, 337, 7

Feldmeier, A., Norman, C., Pauldrach, A. Owocki, S., Puls, J., 
\& Kaper, L. 1997, in 
Mass Ejection From AGN, eds. R. Weymann, I. Shlosman, and N. Arav, 
ASP Conf. Series, 128, 258

Ferland, G.J., Korista, K.T., Verner, D.A., Ferguson, J.W., Kingdon, J.B., 
\& Verner, E.M. 1998, \pasp, 110, 761

Foltz, C.B, Weymann, R.J., Peterson, B.M., Sun, L., Malkan, M.A., \& 
Chaffee, F.H. 1986, \apj, 307, 504

Galavis, M.E., Mendoza, C. \& Zeippen, C.J. 1998, \aaps, 131, 499

Gallagher, S.C., Brandt, W.N., Sambruna, R.M., Mathur, S., \& 
Yamasaki, N. 1999, \apj, 519, 549

Green, P., \& Mathur, S. 1996, \apj, 462, 637

Grevesse, N., \& Anders, E. 1989, in Cosmic Abundances of Matter, 
ed. C.I. Waddington, AIP Conf. Proc., 183, 1

Haehnelt, M.G., Natarajan, P., \& Rees, M.J. 1998, \mnras, 300, 817

Halpern, J.P., Eracleous, M., Filippenko, A.V., \& Chen, K. 1996, \apj, 
464, 704

Hamann, F. 1997, \apjs, 109, 279

Hamann, F., Barlow, T. A., 
Junkkarinen, V. T., \& Burbidge, E. M. 1997a, \apj, 478, 80

Hamann, F., Barlow, T. A., \& 
Junkkarinen, V. T. 1997b, \apj, 478, 87

Hamann, F., \& Brandt, W.N. 2000, \pasp, review in prep.

Hamann, F., \& Ferland, G.J. 1999, \araa, 37, 487

Hamann, F., \& Korista, K.T. 1996, \apj, 464, 158

Hamann, F., Korista, K.T., \& Morris, S.L. 1993, \apj, 415, 541

Hamann, F., Netzer, H., \& Shields, J.C. 2000, \apj, 536, 101

Heckman, T.M., Armus, L., \& Miley, G.K. 1990, \apjs, 74, 833

Heckman, T.M., Dahlem, M., Stephan, Eales, S.A., Fabbiano, G., 
\& Weaver, K. 1996, \apj, 457, 616

Heckman, T.M., Lehnert, M.D., Strickland, D.K., \& Armus, L. 2000, 
\apj, in press

Hewitt, D. \& Burbidge, G. 1993, 
\apjs, 87, 451

Junkkarinen, V., Hewitt, D. \& Burbidge, G. 1991, 
\apjs, 77, 203

Kaspi, S., Smith, P.S., Netzer, H., Maoz, D., Jannuzi, B.T., \& Giveon, 
U. 2000, \apj, 533, 631

Laor, A., Fiore, F., Elvis, M., Wilkes, 
B. J., \& McDowell, J. C. 1997, \apj, 477, 93

Marziani, P., Sulentic, J.W., Dultzin-Hacyan, D., Calvani, M., 
\& Moles, M. 1996, \apjs, 104, 37

Mendoza, C. 1983, in Planetary Nebulae, ed. D.R. Flower, 
(Dordrecht: Reidel), 143

Meurer, G.R., Heckman, T.M., \& Calzetti, D. 1999, \apj, 521, 64

Morris, S.L., Weymann, R.J., Foltz, 
C.B., Turnshek, D.A., Shectman, S., Price, C., \& Boroson, T.A. 1986, 
\apj, 310, 40

Morrison, R., \& McCammon, D. 1983, \apj, 270, 119

Murray, N., \& Chiang, J. 1995, 
\apj, 454, L105

Murray, N., Chiang, J., Grossman, S. A., 
\& Voit, G. M. 1995, \apj, 451, 498

Netzer, H. 1992, AIP Conf. Proc. 254, Testing the AGN Paradigm, ed. S.S. Holt, 
S.G. Neff, \& C.M. Urry (New York:AIP), 146

O'Brien, P.T., Gondhalekar, P.M., \& Wilson, R. 1988, \mnras, 233, 801

Osterbrock, D.E. 1989, Astrophysics of Gaseuos Nebulae and Active 
Galactic Nuclei, Mill Valley, Univ. Sci. Press

Peterson, B.M. 1993, \pasp, 105, 207

Petitjean, P., Rauch, M., \& 
Carswell, R. F. 1994, \aap, 291, 29

Petitjean, P. \& Srianand, R. 1999, \aa, 345, 73

Rauch, M. 1998, \araa, 36, 267

Richards, G.T. 2000, \apjs, submitted

Richards, G.T., Laurent-Muehleisen, R.H. Becker, \& York, D.G. 2000, 
\apj, submitted

Sargent, W.L.W., Boksenberg, A. \& Young, P. 1982, \apj, 252,  54

Savage, B.D., \& Sembach, K.R. 1991, \apj, 379, 245

Stockton, A.N., \& Lynds, C.R. 1966, \apj, 144, 451

Terlevich, R.J., \& Boyle, B.J. 1993, \mnras, 262, 491

Tripp, T.M., Lu, L., \& Savage, B.D. 1996, \apjs, 102, 239

Turnshek, D. A. 1988, in QSO Absorption Lines: 
Probing the Universe, eds. J. C. Blades, D. A. Turnshek, \& C. A. 
Norman (Cambridge: Cambridge Univ. Press), 17

Tytler, D., \& Fan, X.-M. 1992, \apjs, 79, 1

Verner, D., Verner, E.M., \& Ferland, G.J. 1996, Atomic Data and Nucl. 
Data Tables, 64, 1

Vogt, S.S., et al. 1994, Proc. SPIE, 2198, 362

Wampler, E.J., Chigai, N.N., \& Petitjean, P. 1995, \apj, 443, 586

Weymann, R.J., Morris, S.L., Foltz, C.B., \& Chaffee, F.H. 1991, \apj, 
373, 465

Weymann, R.J., Turnshek, D.A., \& Christiansen, W.A. 1985, in Astrophysics 
of Active Galactic Nuclei and Quasistellar Objects, ed. J. Miller 
(Oxford:Oxford Univ. Press), 333

Weymann, R.J., Williams, R.E., Peterson, B.M., \& Turnshek, D.A. 1979, 
\apj, 234, 33

Wilkes, B.J., Tananbaum, H., Worrall, D.M., Avni, Y., Oey, M.S., 
\& Flanagan, J. 1994, ApJS, 92, 53

Williams, R.E., Strittmatter, P.A., Carswell, R.F., \& Craine, E.R. 
1975, \apj, 202, 296

Wills, B.J., Thompson, K.L., Han, M., 
Netzer, H., Wills, D. \etal 1995, \apj, 447, 139

Zheng, W., Kriss, G.A., Telfer, R.C., Grimes, J.P., \& Davidson, 
A. F. 1997, \apj, 475, 469

\bigskip\bigskip

\section*{FIGURE CAPTIONS}
\parskip=0pt
\leftskip=0pt
\parindent=1.5em

\bigskip

Figure 1. --- Part of the measured spectrum of 3C 191 showing 
the strong associated absorption lines (labeled above). 
The flux has units 10$^{-15}$ ergs cm$^{-2}$ s$^{-1}$ \AA$^{-1}$.
\medskip

Figures 2a and 2b. --- AAL profiles on a 
velocity scale defined by the BEL redshift, $z_e = 1.956$. The lines 
\ion{C}{2} \lam 1335 and \ion{C}{2}$^*$ \lam 1336 
are blended togther, as are \ion{C}{4} \lam 1548 and 
\lam 1551, causing them to appear in each other's panel. 
The feature labeled ``\ion{N}{5} average'' is an 
average of the \ion{N}{5} doublet \lam 1239 and 
\lam 1243. The sharp features overlying the \ion{Fe}{2} \lam 2587 and 
\lam 2600 profiles are telluric O$_2$ lines. Dotted vertical 
lines are drawn to guide the eye. 
\medskip

Figure 3. --- Observed profiles of \ion{Mg}{2} \lam 2804, 
\ion{Al}{3} \lam 1863 and \ion{Fe}{2} 
\lam 2344 (bold solid lines in each panel) are compared to predicted 
profiles in these lines based on \ion{Mg}{2} \lam 2796, 
\ion{Al}{3} \lam 1855 
and \ion{Fe}{2} \lam 2383 (thin lines). 
The thin curve near the bottom of each panel shows the 
variance per measured pixel. The vertical dotted lines mark the 
same velocities as Figure 2. 
The deeper observed compared to predicted profiles 
indicate partial coverage of the background light source. See 
\S4.2.
\medskip

Figure 4. --- 
Theoretical ionization fractions in photoionized, optically 
thin ($\tau_{LC}<1$) clouds having different ionization parameters, $U$. 
The curves for different metal ions are labeled near their peak values 
whenever possible. 
The curves for \ion{Fe}{2}, \ion{Al}{2} and \ion{Al}{3} are dashed 
for clarity. The \ion{H}{1} fraction is shown across the top. See 
\S4.4.
\medskip

Figure 5. --- 
Theoretical column density ratios in photoionized, 
optically thin clouds with solar element abundances and different $U$. 
The measured column densties (Table 1) compared favorably with these 
predictions if $-3.4 \la\ \log U\ \la\ -2.8$. See \S4.7.
\medskip

Figure 6. --- 
Schematic representation of the AAL environment, 
showing pockets of dense neutral gas (filled black circles) 
surrounded by a less dense and more highly ionized medium (grey 
circles). The more extended regions produce smoother AAL 
profiles and more complete spatial coverage of the background 
light source (\S5.2). 

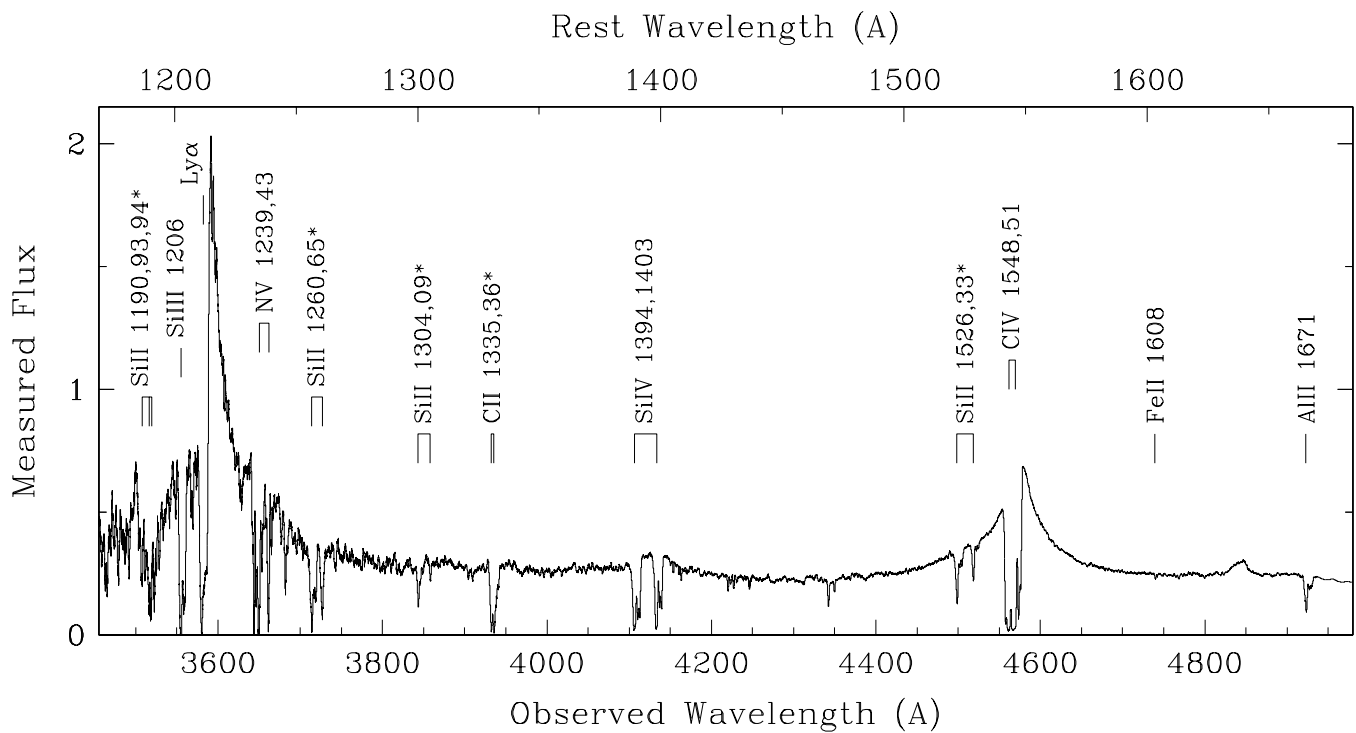
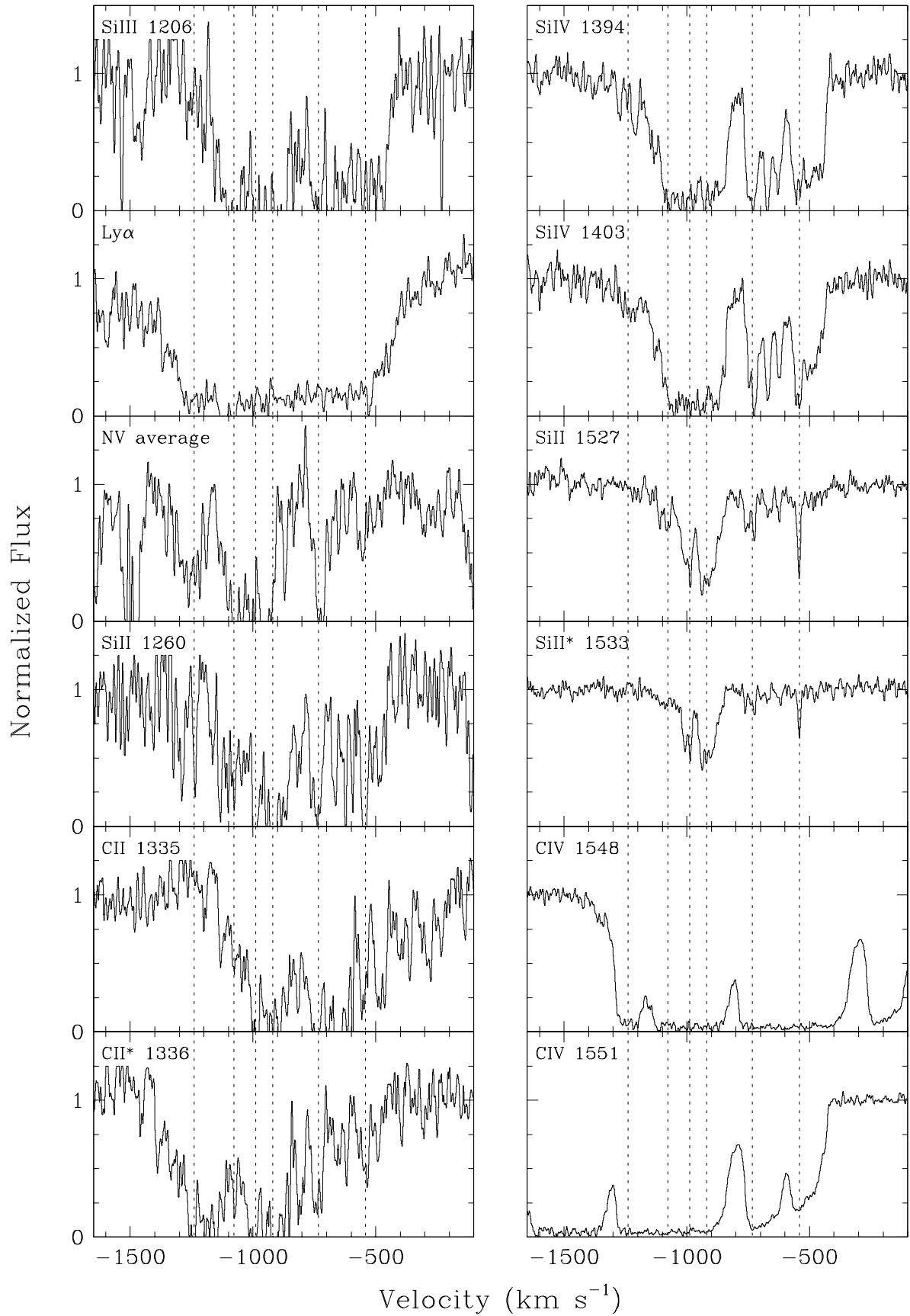
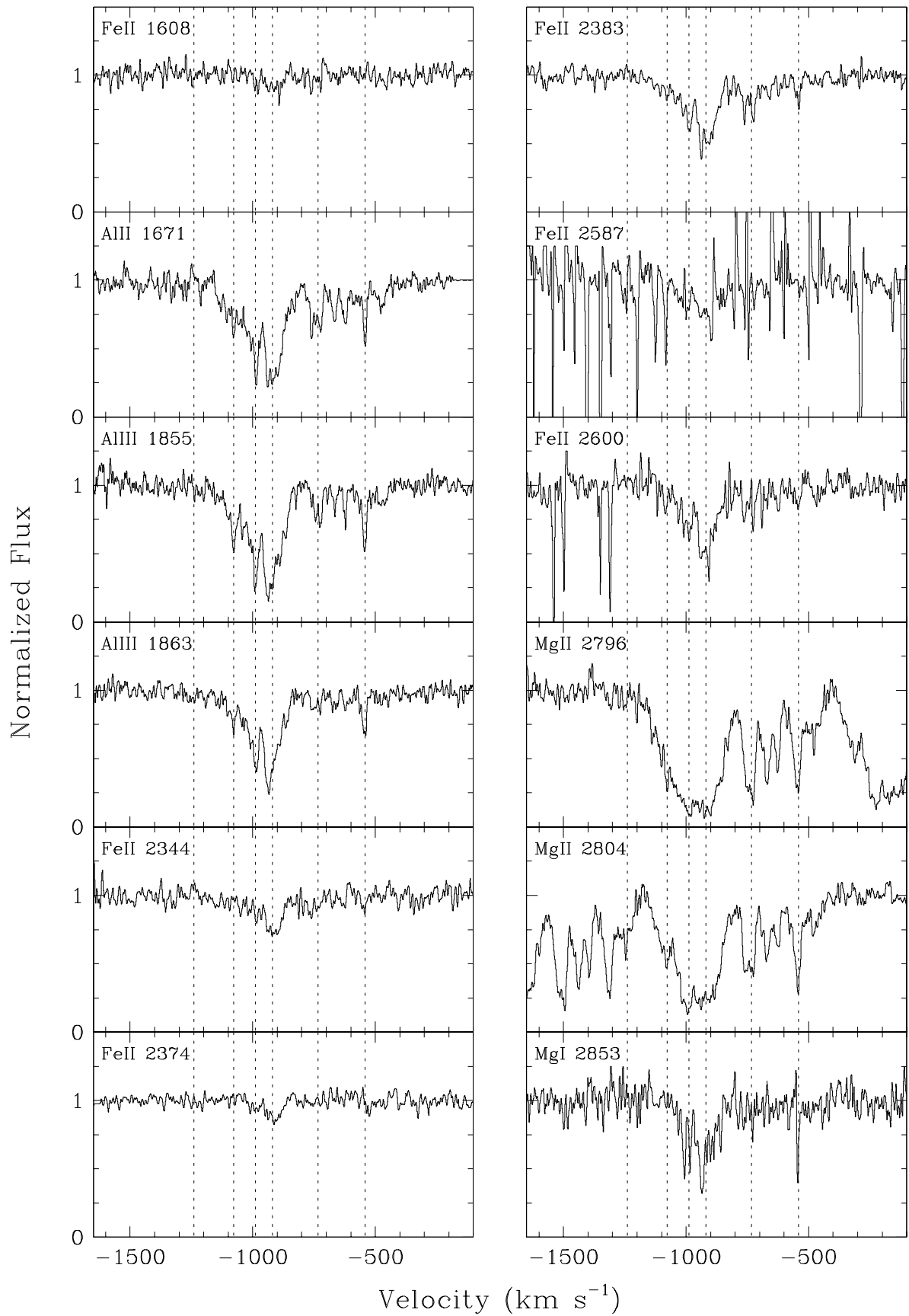
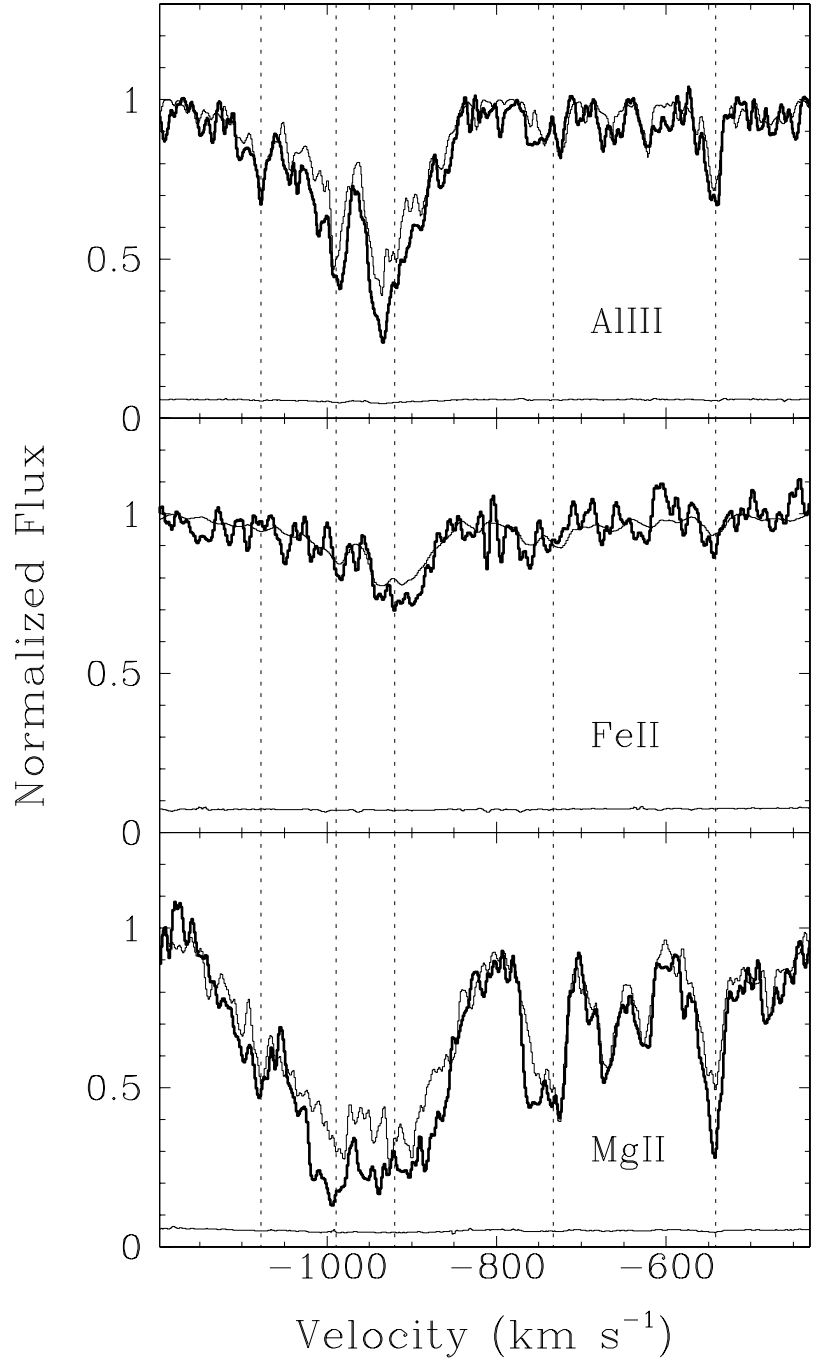
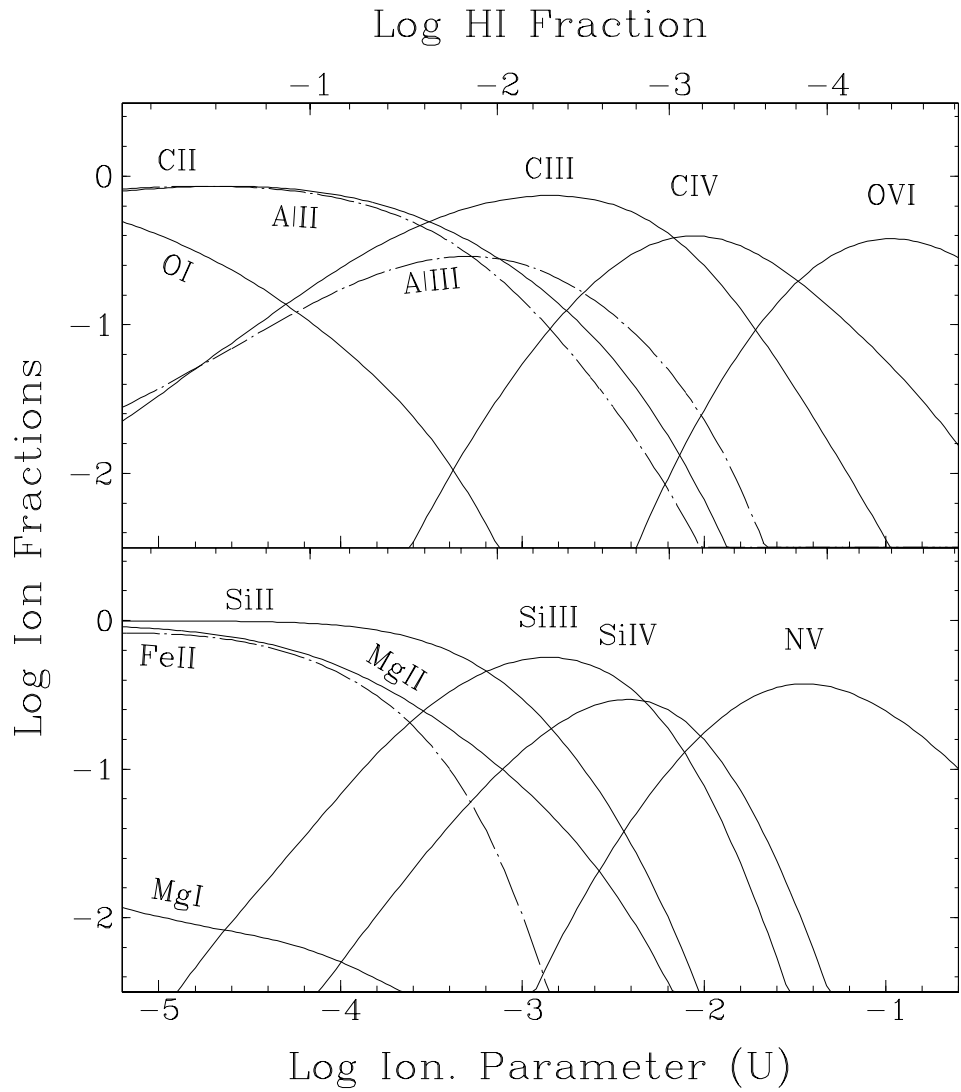
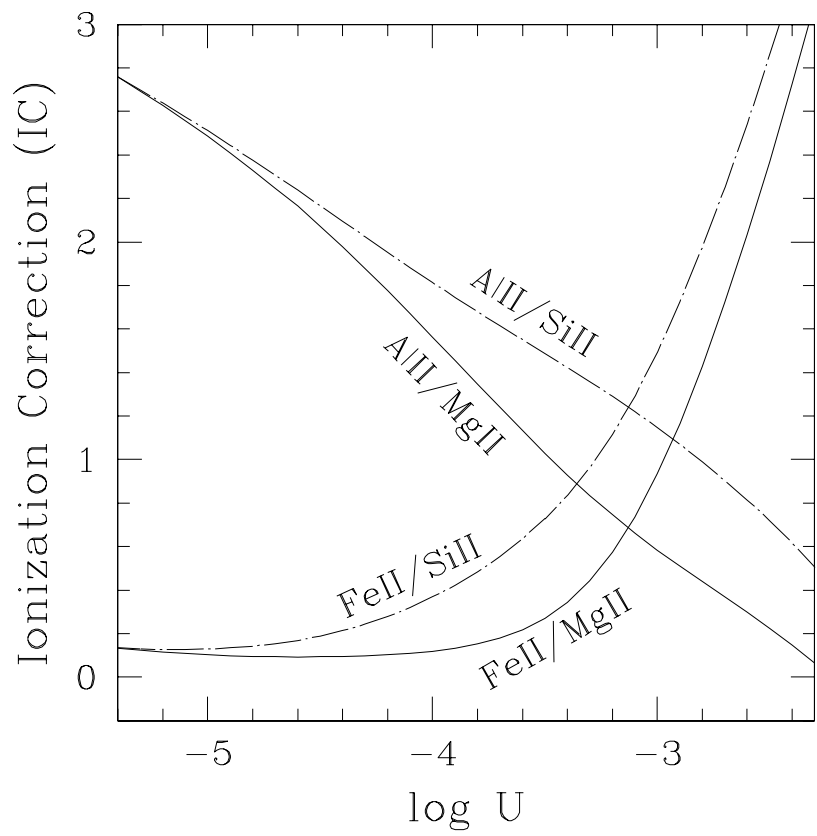
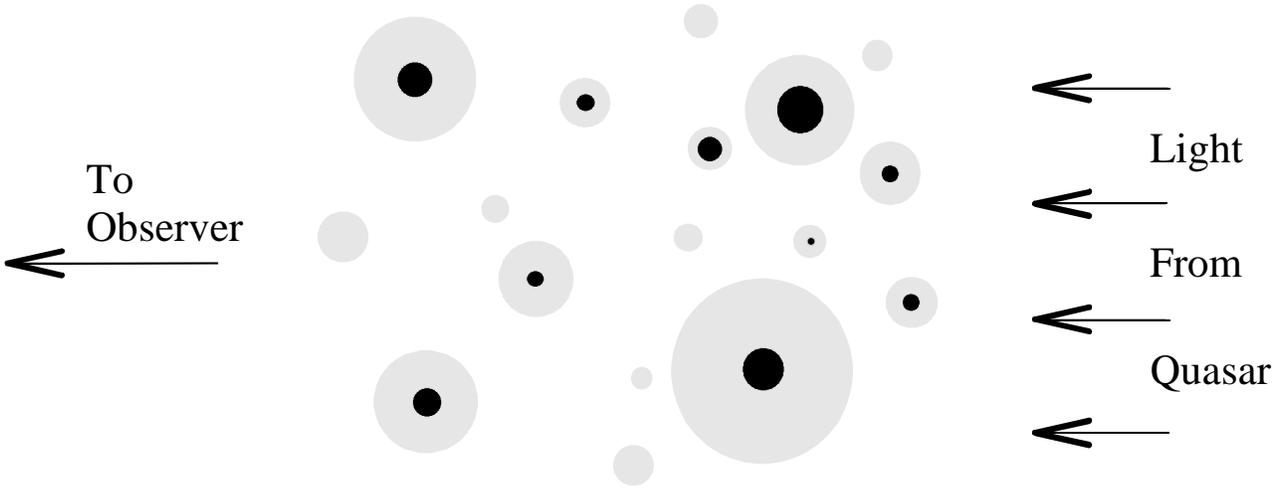
\end{document}

%% file: table1.tex
\begin{deluxetable}{lcccccccccc}
\tablewidth{0pt}
\tablecaption{AAL Measurements}
\tablehead{\colhead{}& \multicolumn{4}{c}{----------- W$_{\lambda}$(\AA ) 
-----------}& \colhead{~~}& \multicolumn{5}{c}{------------------ $\log N$(cm$^{-2}$) 
------------------}\\\colhead{Line~~~~~~~~~~~}& 
\colhead{$\Delta v_1$}& \colhead{$\Delta v_2$}& \colhead{$\Delta v_3$}& 
\colhead{Total}& \colhead{~}&
\colhead{$\Delta v_1$}& \colhead{$\Delta v_2$}& \colhead{$\Delta v_3$}& 
\colhead{Total}& \colhead{Tot($C_f<1$)\tablenotemark{a}}}
\startdata
\ion{Si}{3} 1206&     0.04& 1.46& 1.19& 2.70& &   12.5&  14.5&  14.2&  14.6& ---\\
\Lya&                 0.69& 1.29& 1.28& 3.26& &   14.4&  14.9&  14.7&  15.2& ---\\
\ion{N}{5} average&   0.36& 0.96& 0.38& 1.70& &   14.5&  15.4&  14.8&  15.5& ---\\
\ion{Si}{2} 1260&     0.12& 1.04& 0.80& 1.95& &   13.0&  14.3&  14.0&  14.5& ---\\
\ion{Si}{2}$^*$ 1265&  ---& 0.85& 0.14& 0.97& &    ---&  14.1&  13.1&  14.2& ---\\
\ion{Si}{2} 1304&      ---& 0.44& 0.19& 0.63& &    ---&  14.7&  14.3&  14.8& ---\\
\ion{Si}{2}$^*$ 1309&  ---& 0.13& 0.04& 0.17& &    ---&  14.2&  13.6&  14.3& ---\\
\ion{C}{2}+\ion{C}{2}$^*$&      ---&  ---&  ---& 2.49& &    ---&   ---&  
---& 15.2& ---\\
\ion{Si}{4} 1394&     0.18& 1.29& 1.17& 2.63& &   13.4&  14.6&  14.4&  14.8& ---\\
\ion{Si}{4} 1403&     0.14& 1.22& 0.97& 2.32& &   13.5&  14.8&  14.6&  15.0& ---\\ 
\\
\ion{Si}{2} 1527&     0.03& 0.65& 0.32& 1.00& &   13.0&  14.5&  14.1&  14.7& 15.1\\
\ion{Si}{2}$^*$ 1533&  ---& 0.38& 0.08& 0.46& &    ---&  14.2&  13.5&  14.3& 14.5\\
\ion{C}{4} 1548&      0.71& 1.66&  ---&  ---& &   14.6&  15.1&   ---&   ---& ---\\
\ion{C}{4} 1551&       ---& 1.66& 1.39&  ---& &    ---&  15.4&  15.2&   ---& ---\\
\ion{C}{4} total&      ---&  ---&  ---& 6.12& &    ---&   ---&   ---&  15.7& ---\\
\ion{Fe}{2} 1608&      ---& 0.04\rlap{:}&  ---& 0.04\rlap{:}& &    ---&  
13.4\rlap{:}&   ---&  13.4\rlap{:}& 13.5\rlap{:}\\
\ion{Al}{2} 1671&      ---& 0.72& 0.37& 1.09& &    ---&  13.4&  13.0&  13.5& 13.9\\
\ion{Si}{2} 1808&      ---& 0.05\rlap{:}&  ---& 0.05\rlap{:}& &    ---&  
14.9\rlap{:}&   ---&  14.9\rlap{:}& 15.0\rlap{:}\\
\ion{Al}{3} 1855&      ---& 0.77& 0.25& 1.02& &    ---&  13.8&  13.2&  13.9& 14.4\\
\ion{Al}{3} 1863&      ---& 0.60& 0.19& 0.78& &    ---&  14.0&  13.4&  14.1& 14.3\\
\\
\ion{Fe}{2} 2344&      ---& 0.30& 0.08& 0.38& &    ---&  13.7&  13.2&  13.9& 14.0\\
\ion{Fe}{2} 2374&      ---& 0.08\rlap{:}&  ---& 0.08\rlap{:}& &    ---&  
13.5\rlap{:}&   ---&  13.5\rlap{:}& 13.7\rlap{:}\\
\ion{Fe}{2} 2383&      ---& 0.63& 0.30& 0.92& &    ---&  13.6&  13.3&  13.8& 14.0\\
\ion{Fe}{2} 2587&      ---& 0.30\rlap{:}&  ---& 0.30\rlap{:}& &    ---&  
14.0\rlap{:}&   ---&  14.0\rlap{:}& 14.2\rlap{:}\\
\ion{Fe}{2} 2600&      ---& 0.59& 0.23& 0.82& &    ---& 
13.7&  13.2&  13.9& 14.1\\
\ion{Mg}{2} 2796&     0.07& 2.10& 1.39& 3.55& &   12.3&  14.0&  13.7&  14.1& 14.4\\
\ion{Mg}{2} 2804&      ---& 1.69& 1.00& 2.70& &    ---&  14.1&  13.8&  14.3& 14.4\\
\ion{Mg}{1} 2853&      ---& 0.55& 0.24& 0.79& &    ---&  12.7&  12.3&  12.9& 13.1\\
\enddata
\tablenotetext{a}{The total column densities in the last column are corrected for 
$C_f \approx 0.85$ in the \ion{Mg}{2} doublet and $C_f \approx 0.70$ in 
other lines. Lines without entries in this column are poorly measured or 
appear very optically thick (\S4.3).}
\end{deluxetable}